\documentclass[a4paper]{article}
\usepackage{graphicx}

\usepackage{amsmath,amssymb}
\newcommand{\ba}{\begin{array}}
\newcommand{\ea}{\end{array}}
\newcommand{\bea}{\begin{eqnarray}}
\newcommand{\eea}{\end{eqnarray}}
\newcommand{\be}{\begin{equation}}
\newcommand{\ee}{\end{equation}}

\newcommand{\ve}[1]{\mathbf{#1}}
\newcommand{\re}[1]{(\ref{#1})}

\def\trans{\mbox{\tiny$\bot$}} % Transverse component
\def\longi{\mbox{\tiny$\|$}}   % Longitudinal component

\begin{document}

\begin{center}

{\bf PULSATIONS OF THE ELECTRON-POSITRON PLASMA IN THE FIELD \\
OF OPTICAL LASERS}

\vspace{4mm} D.B. Blaschke$^{1,2,3}$, A.V. Prozorkevich$^4$,
S.A. Smolyansky$^{3,4}$ and A.V.~Tarakanov$^4$ \\
[15pt]

\textit{$^1$\it Fakult\"at f\"ur Physik, Universit\"at
Bielefeld, D-33615 Bielefeld, Germany} \\
\textit{$^2$\it Joint Institute for Nuclear Research, 141980, Dubna, Russia}\\
\textit{$^3$\it Fachbereich Physik, Universit\"at Rostock,
Rostock, D-18051, Germany}\\
\textit{$^4$Physics Department, Saratov State
University, 410026, Saratov, Russia} \\[5pt]

\end{center}

\begin{abstract}
The possibility to observe vacuum electron-positron pair
creation due to a powerful optical laser pulse is discussed.
We employ a quantum kinetic formulation of the problem with
a source term describing the vacuum pair production in a
homogeneous electric field with arbitrary time dependence
(dynamical Schwinger mechanism).
For a periodic field weak in comparison with
the critical value $E_{cr}=m^2/|e|$, the electron-positron
plasma density changes approximately periodically with
twice the field frequency.
Under these conditions, the mean value $<n>$ for the density
per period in the volume $\lambda^3$ is a more appropriate
characteristic quantity than the residual density
$n_r$ which is taken over an integer number of field periods
and calculated using the imaginary time method.
The value $<n>$ is proportional to the squared field intensity
and does not depend on the frequency. We show that in terms
of the parameter $<n>$ an optical laser can be more effective
than a X-ray one. We expect that it is possible
to observe the vacuum creation effect not only by means of
the planned X-ray free electron lasers but already at present-day
optical lasers.
\end{abstract}

%\keywords{pair creation, optical laser, free electron lasers, kinetic
%equation, electron-positron plasma, mean pair density}

\section{INTRODUCTION}

QED is considered as the most advanced physical theory, many
of its predictions have been proven experimentally with highest
available precision.
Nevertheless, some questions are discussed till now, e.g.,
the vacuum pair creation effect by a classical electric field
\cite{Schwinger}.
A complete theoretical description of this effect has been obtained
\cite{Grib,Nick,Greiner,Fradkin}, but there is still no
experimental proof.
The main problem is the high value of the critical electric field strength,
necessary to be reached for the pair creation,
namely $E_{cr} = 1.3\times 10^{16} V/cm$ for electron-positron pairs.
According to the Schwinger formula, the pair creation rate in a constant
electric field is
\begin{equation}\label{fs}
 \frac{dN}{d^3x dt}=\frac{(eE)^2}{4\pi^3}\sum\limits_{n=1}^{\infty}
 \frac{1}{n^2}\exp{\left(-  n \pi \frac{ E_{cr}}{E}\right)}
\end{equation}
and therefore exponentially suppressed when $E\ll E_{cr}$.
Fortunately, the situation changes qualitatively if the field acts
a finite time only \cite{Grib,Smol,Rob,Pop}. In this case,
the Schwinger formula as well as its analog for a
monochromatic field (Brezin-Itzykson formula \cite{Brezin})
become inapplicable.

There are a few examples for physical situations where the
Schwinger effect can be observed, e.g. relativistic heavy ion
collisions \cite{rhic}, neutron stars \cite{magn,Ruffini} and
focussed laser pulses \cite{focus}. The structure of a real laser
field is too complicated for the analysis, because the Schwinger
effect is non-perturbative and it requires the exact solution of
the dynamical equations. That is why the approximation of the
homogeneous electric field is used in most cases. According to
different estimates \cite{Pop,Brezin,Bunkin,Marinov} the effect of
vacuum creation can not be observed with the presently achieved
level of laser power, see also \cite{BulanovSS}.

The recent development of laser technology, in particular the
invention of the chirped pulse amplification method, has resulted
in a huge increase of the light intensity in a laser focal spot
\cite{Mou,BulanovSV}. The most advanced lasers produce pulses with
intensities reaching $10^{22}$ W/cm$^2$ and the pulse duration
decreasing down to few oscillation periods. As the construction of
X-ray free electron lasers {XFELs} \cite{Ring} is now planned, the
possibility of the experimental proof of the Schwinger effect
attracts attention again. The non-stationary effects become
important under conditions of short pulses. We use in our work the
kinetic equation approach, which allows us to consider the
dynamics of the creation process taking into account  the initial
conditions \cite{Smol}. Compared to the other treatments, the
approach within the framework of a transport equation contains
some new dynamical aspects, such as longitudinal momentum
dependence of the distribution functions and non-markovian
character of the time evolution. It takes into account the effects
of the field switching and statistics, as well \cite{our}. This
approach has been applied already to the periodical field  case
 \cite{Rob} with near-critical values of the field strength and
X-ray frequencies. In particular, it was shown that there
is an "accumulation" effect when the intensity of the field
is about half critical: the average density of pairs
grows steadily with the increase of the field period
numbers.
%That pair creation can occur in a cold collisionless plasma, 
%under action of an electromagnetic field which is an actual solution of
%the Maxwell equations was discovered in the works
%\cite{BulanovSS}. It was shown, particulary, that 
%already for intensities of the laser field smaller than the critical 
%value, has been discussed in the works \cite{BulanovSS}.
%the energy loss due to pair creation is the same order of the energy
%storied in the laser pulse. Therefore  the backreaction of the
%produced pair on the laser field should be taken into account.}

In the present work, we consider the other region of field
parameters really achievable nowadays in the optical lasers: $E\ll
E_{cr}$ and $\nu\ll m$, where $\nu$ is the laser field frequency.
We suggest to use in the criterion for the creation efficiency
the mean value $<n>$ for the density
per period in the volume $\lambda^3$ is a more appropriate
characteristic quantity than the residual density
$n_r$. The latter is taken over an integer number of field periods
and calculated using the imaginary time method.
The main result is that optical lasers can generate a
greater density $<n>$ than X-ray ones.

The work is organized as follows. Section 2 contains  the
statement of the problem and the necessary information
about the kinetic equation which is used for the description of
vacuum pair creation. We  solve this equation numerically
for the conditions of the SLAC experiment \cite{Bula} and study
some features of pair production dynamics. We compare here
our results obtained on the non-perturbative basis with the
predictions of another approach \cite{Pop} and show that
optical lasers can be effective generators of
electron-positron pairs during the action of a laser pulse.
In Section 3, the low density approximation is considered.
It allows to get some analytical results and to make simple estimates.
Finally, in Section 4, we discuss some possibilities of direct
experimental verification of pair production by high power optical
lasers. 
%We summarize our results and present the conclusion.

\section{THE KINETIC EQUATION APPROACH}

In the kinetic approach \cite{Smol}, the basic quantity is
the distribution function of electrons in the momentum
representation $f(\ve{p},t)$. The kinetic equation  for
this function is derived from the Dirac equation in an
external time-dependent field by the canonical Bogoliubov
transformation method \cite{Grib}, or by the help of the
oscillator representation \cite{OR}. This procedure is
exact but valid only for the simplest field configurations,
e.g., the homogeneous time dependent electric field with
the fixed direction
\begin{equation}\label{field}
\mathbf{E}(t)=(0,0,E(t)),\qquad E(t)=-\dot A(t)~,
\end{equation}
where the vector potential is given in the Hamiltonian gauge
$A^\mu=(0,0,0,A(t))$ and the overdot denotes the time
derivative. Such a field is not appropriate for a quantitative
description of the laser pulse, but can probably be used as
qualitative model to estimate results.
The corresponding kinetic equation in
the collisionless limit has the form \cite{Smol}
\begin{equation}\label{ke}
\frac{df(\ve{p},t)}{dt} = \frac12 \Delta(\ve{p},t)\int\limits_{t_0}^t \! dt'
\, \Delta(\ve{p},t')\left[ 1-2f(\ve{p},t')\right]\cos{ \theta(\ve{p},t',t)},
\end{equation}
where
\begin{eqnarray}
\Delta(\ve{p},t)&=&eE(t)\frac{\sqrt{m^2+p_{\trans}^2}}{\omega^2(\ve{p},t)},\\
\omega(\ve{p},t)&=&\sqrt{m^2+p_{\trans}^2+[p_{\longi}-eA(t)]^2},\\
\theta(\ve{p},t',t)&=&2\int\limits_{t'}^{t}dt_1\,\omega(\ve{p},t_1).
\end{eqnarray}
Eq. \re{ke} can be transformed to a system of ordinary
differential equations, which is convenient for a numerical analysis
\begin{eqnarray}\label{ode}
  \dot{f} &=& \frac{1}{2}\, \Delta v_1, \nonumber \\
  \dot{v}_1 &=& \Delta (1 - 2f)- 2 \omega v_2, \nonumber\\
  \dot{v}_2 &=& 2 \omega\, v_1,
\end{eqnarray}
where $v_1, v_2$ are real auxiliary functions. The system \re{ode} is
integrated via the Runge-Kutta method with the initial conditions
$f(\ve{p},t_0)=v_1(\ve{p},t_0)=v_2(\ve{p},t_0)=0$.
The momentum dependence of the distribution function is defined by means
of a discretization of the momentum space in a 2-dimensional grid, where
the system \re{ode} is solved in each of its nodes.
The concrete grid parameters depend on the field strength, where
typical values are $\Delta p\approx 0.05~m$ (grid step) and $p_{max}\approx
(5-10)~m$ (grid boundary). The particle number density can be found after
that as a moment of the distribution function
\begin{equation}\label{dens}
n(t)=2 \int\frac{d^3p}{(2\pi)^3} f(\ve{p},t)\ .
\end{equation}
Let us consider a harmonic time dependence of the field
\begin{equation}\label{harm}
E(t)=E_m \sin{\nu t}, \qquad A=-\frac{E_m}{\nu} \cos{\nu t}.
\end{equation}
The time dependence of density for the field \re{harm} with the
parameters $E_m/E_{cr}= 4.6\cdot 10^{-6}$ and $\nu/m = 4.29\cdot
10^{-5}$, corresponding to SLAC experiments \cite{Bula} is shown
in Fig. 1 in comparison to the planned X-ray laser \cite{Ring}
with $E/E_{cr}=0.24$ and $\nu/m = 0.0226$. The pair density
oscillates with twice the frequency of the laser field.
The density value $n_r$, which is evaluated in the imaginary time method
\cite{Pop,Marinov}, corresponds to an integer number of field
periods, $n_r=n(t=2\pi/\nu)$, and it is negligible in comparison
with the density value $n_m$ corresponding to the electric field
maximum, $n_m=n(t=\pi/2\nu)$. The mean density per period $<n>$ is of
the same order as $n_m$. For the conditions of the SLAC experiment
the ratio of $<n>/n_r \approx 3\cdot 10^{11}$. As a consequence, in
spite of the fact that the residual density for the X-ray laser
exceeds the one for the optical laser by a large factor, the situation
is different regarding the mean density: the optical laser can produce
more pairs per volume lambda cubed than the X-ray one.
\begin{figure}[t]
\centerline{
\includegraphics[width=120mm,height=70mm]{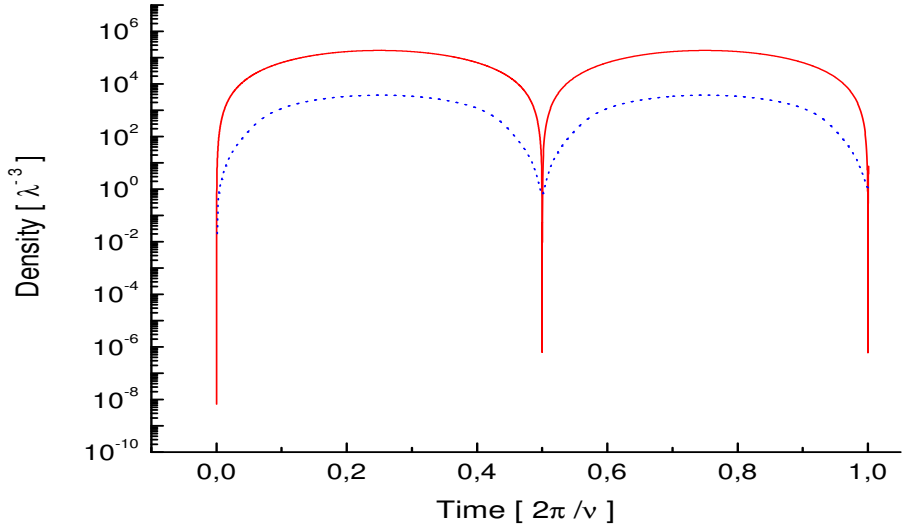}}
\caption{Time dependence of the density $n(t)$ in the
volume $\lambda^3$ in the weak periodic field with the
parameters $E/E_{cr}=4.6\cdot 10^{-6}$ and $\nu/m =
4.29\cdot 10^{-6}$, corresponding to the SLAC experiment
\cite{Bula} (solid line) and in the near-critical field case
of an X-ray laser \cite{Ring} with $E/E_{cr}=0.24$ and $\nu/m =
0.0226$ (dashed line).}
\end{figure}

\section{LOW DENSITY APPROXIMATION}

The low density approximation $f\ll 1$ can be used in the
weak field limit $E\ll E_{cr}$. In that case it is possible to obtain
analytic estimates for the residual ($n_r$) and the maximal ($n_m$)
densities the Eq. \re{ke}.

\begin{figure}
\centering
\includegraphics[width=70mm,height=65mm]{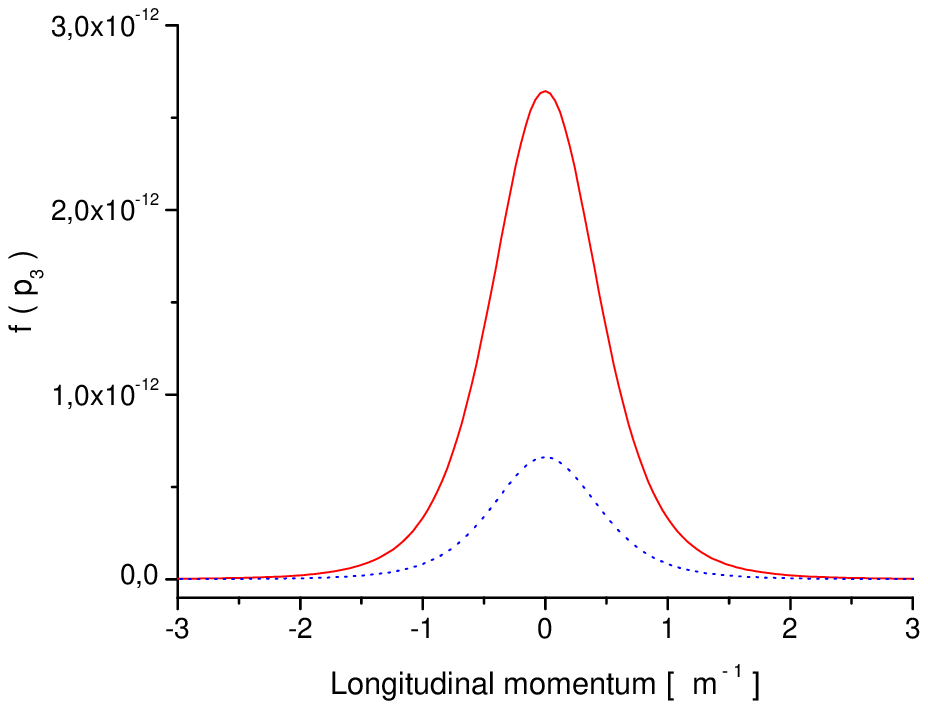}
\includegraphics[width=70mm,height=65mm]{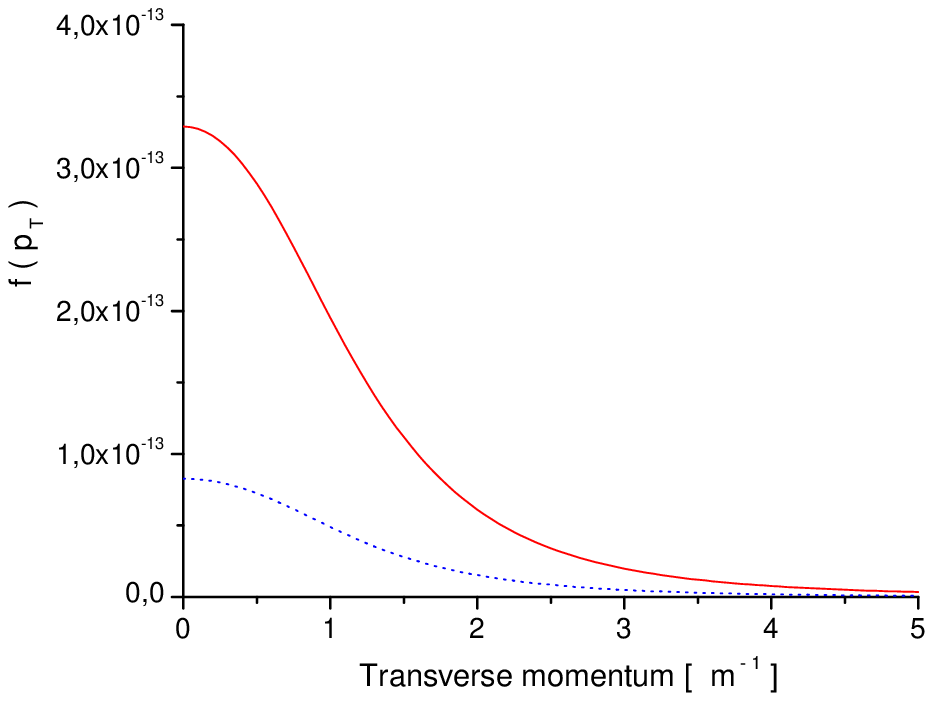}
\caption{The transverse (left panel) and the longitudinal (right panel)
momentum dependence of the distribution function at $t=\pi/2\nu$
(maximum of the field) for the two values of the field strength:
$E_m = 4.6\cdot 10^{-6}E_{cr}$ (solid lines) and $E_m = 3\cdot
10^{-6}E_{cr}$ (dashed lines) with optical frequency $\nu =
4.26\cdot 10^{-5}~m$.}\label{spectr}
\end{figure}

The particle density in this approximation is
\begin{eqnarray}
n(t)=\frac{e^2}{(2\pi)^3}\int d\ve{p} \,\varepsilon_\perp^2\,
\int\limits_{t_0}^t dt_1 \, \frac{E(t_1)}{\omega^2(t_1)}
\int\limits_{t_0}^{t_1} dt_2 \, \frac{E(t_2)}{\omega^2(t_2)}
\cos{\left(2\int\limits_{t_2}^{t_1} dt_3 \omega (t_3)\right)}
\end{eqnarray}
and can be transformed to \cite{spie}
\begin{eqnarray}\label{fact}
n(t)=\frac{1}{2(2\pi)^3}\int d\ve{p} \,\varepsilon_\perp^2\, \left|\,
\int\limits_{t_0}^t dt_1 \, \frac{eE(t_1)}{\omega^2(t_1)}
\exp{\left(2i\int\limits_{t_1}^{t} dt_3 \omega (t_3)\right)}\right|^{\,2}
\,.
\end{eqnarray}
Let us assume additionally that the condition
\begin{equation}\label{anz2}
\gamma = \frac{m\,\nu}{|e|E_m}  \gg 1,
\end{equation}
is satisfied, where $\gamma$ is the adiabaticity parameter
\cite{Pop}.  This relation can be treated as the condition for
quasi-classical charge transport in an external field on the time
scale $\sim 1/\nu$, if only the pairs are created with vanishing
momenta.
The latter condition was often used in relation to the longitudinal
momentum \cite{Casher} but the real momentum distribution of
the electron-positron pairs has a width of the order of the inverse
mass for both transverse and longitudinal momenta, see Fig. \ref{spectr}.
The momentum distribution shape varies essentially at the moments
of time corresponding to the field minima: a complex quasi-periodic
structure with a mean period of about the inverse laser frequency is
formed. The mean period value of such a structure decreases
proportionally to the number of field periods.

By means of the inequality \re{anz2} Eq. \re{fact} is reduced to
\begin{equation}\label{fact1}
n(t)=\frac{1}{2(2\pi)^3}\int d\ve{p} \,
\frac{\varepsilon_\perp^2}{\omega^4}\,
\left|\,\int\limits_{t_0}^t dt_1 \, eE(t_1) \exp(2i \omega
t_1)\, \right|^{\,2}.
\end{equation}
The time integral is calculated analytically for the field
\re{harm} and $t_0=0$ with the result
\begin{multline}\label{13}
n(t)=\frac{1}{2(2\pi)^3}\int d\ve{p} \,
\frac{\varepsilon_\perp^2}{\omega^4}\, \left(\frac{eE_m
}{\nu^2-4\omega^2}\right)^2 \biggl\{ \nu^2 (1+\cos^2 \nu t)
+ 4 \omega^2 \sin^2 \nu t  \\ - 2\nu \bigl[ \nu \cos\nu t
\cos 2\omega t +2\omega \sin\nu t \sin 2\omega t
\bigr]\biggr\}.
\end{multline}
According to Eq. \re{13} the residual pair density after $N$ periods
is  $n_r = n(2\pi N/ \nu )$ and mean pair density per period  $<n>$
are estimated as
\begin{eqnarray}\label{15}
n_r &=& \frac{1}{(2\pi)^3}\int d^3p \,
\frac{\varepsilon_\perp^2}{\omega^4}\, \left(\frac{eE_m
\nu}{\nu^2-4\omega^2}\right)^2 \left[1 -
\cos{\left(\frac{4\pi N \omega}{\nu}\right)}\right],  \\
<n> &=&\frac{1}{4(2\pi)^3}\int d^3p \,
\frac{\varepsilon_\perp^2}{\omega^4}\, \left(\frac{eE_m
}{\nu^2-4\omega^2}\right)^2 \bigl[ 3 \nu^2 + 4 \omega^2
\bigr]. \label{16}
\end{eqnarray}
Now we omit the fastly oscillating term in \re{15} and
suppose additionally that %$\nu \ll m$.
\begin{equation}\label{anz3}
    \nu \ll m.
\end{equation}
Then we obtain the simple estimate
\begin{equation}\label{depres}
  \frac{\langle n\rangle}{n_r} \sim \left(\frac{m}{\nu}\right)^2.
\end{equation}
The mean density of electron-positron pairs is defined in this
case  only by the field amplitude and does not depend on the frequency
within a wide range of parameters \re{anz2}. After the integer
period number (when the electric field vanishes) the overwhelming
part of pairs is absorbed and the residual density, which is
estimated within the usual approach \cite{Marinov}, is negligible
in comparison with the mean one used above.
For the Terawatt Nd-glass laser with the wavelength $527$ nm and
the field strength $E_m=6\cdot10^{10}$ V/cm \cite{Bula} we have
$m/\nu \approx 2\cdot 10^5$, so the mean density exceeds the residual
one by more than $10$ orders of magnitude.
According to Fig. 1, there are $\approx 10^{5}$ pairs in a volume
of wavelength cubed on the average for one period of the laser field.
The same pair density can be achieved under the conditions of an
X-ray laser \cite{Rob}.
Let us notice, that the formula from Ref. \cite{Brezin} for the pair creation
probability with the condition \re{anz2}
\begin{equation}\label{Brez}
w\simeq \frac{ (eE)^2}{8}\left( \frac{eE}{2m\omega}
\right)^{4m/\omega}
\end{equation}
gives only a negligible creation probability $\approx
10^{-10^5}$ in this case. This is not surprising because
the formula \re{Brez} is not applicable for field pulses
of finite duration.

\section{SUMMARY}

The simplest laser field model \re{harm} predicts the existence of
a dense electron-positron plasma during a laser pulse
duration, which is absorbed almost completely after switching
off the field. The mean density is defined by the field strength
and does not depend on frequency. The plasma density reaches
$10^{18}$ cm$^{-3}$ within the range of the really achieved fields
$10^{10}-10^{11}$ V/cm. The usual recipe for an experimental
proof of the Schwinger effect suggests that an increase of the
residual plasma density can be achieved by an increase of both
the frequency and the power of the laser radiation, e.g., via
XFEL facilities \cite{Ring}.
It is possible that the simpler way is to try to probe the plasma
with some external field in addition to the generating one.
SLAC experiments \cite{Bula} can serve as an example of this type,
where a high energy ($46$ GeV) electron beam crosses the focus of
the $527$ nm Terawatt pulsed Nd-glass laser and $106\pm 14$ positrons
above background have been observed.
The observed positrons are interpreted as arising from a
two-step process: laser photons are backscattered to GeV energies
by the electron beam and a collision between the high-energy
photon and several laser photons produces an electron-positron
pair.

%In conclusion, we shall mention some possibility of the
%observation of the discussed effect. The motion of the
%electron-positron plasma in the laser pulse spot is
%accompanied by a loss of the fast particles which become
%accessible for registration. On the other hand, there is
%some probability of electron-positron pairs annihilation
%and creation of the $\gamma$ - quantum with the
%characteristic energy $\sim 0.5$ MeV. We plan to make
%appropriate estimates in a subsequent work.

\vspace{2mm} \textbf{Acknowledgments.} This work was
supported partly by Russian Federations State Committee for
Higher Education under grant E02-3.3-210 and Russian Fund
of Basic Research (RFBR) under grant 03-02-16877. S.A.S.
acknowledges support by the DFG Graduiertenkolleg 567
"Stark korrelierte Vielteilchensysteme" at the University
of Rostock and by DFG grant No. 436 RUS 117/78/04.

\end{document}